\def\gcmm3{{\,{\rm g\,cm^{-3}}}}

\def\fun#1#2{\lower3.6pt\vbox{\baselineskip0pt\lineskip.9pt
  \ialign{$\mathsurround=0pt#1\hfil##\hfil$\crcr#2\crcr\sim\crcr}}}
\documentstyle[preprint,prd,aps,epsfig,floats,graphics]{revtex}
\draft

\begin{document}

\tighten

\title{
\vskip1.55cm
Notes on a quantum gravitational collapse}

\author{Victor Berezin}

\address{Institute for Nuclear Research of the Russian Academy of Sciences,
60th October Anniversary Prospect, 7a, 117312, Moscow, Russia\\
e-mail: berezin@ms2.inr.ac.ru}
\maketitle

  Everybody knows what the classical black holes are. In short, this
  is a spacetime region beyond the so-called event horizon. The notion of
  the event horizon is mathematically well defined. The situation with
  a definition of quantum black hole is not so clear. The problem is
  that the classical event horizon can be defined only globally, i.e.
  in order to be sure we have a black hole we would need an infinite time 
  interval. But, in classical physics we have trajectories off all the
  particles and equations of motion for all the fields and can, in
  principle, construct some ideal models for the gravitational collapse 
  and study the black hole formation under different conditions.
In quantum theory we have no trajectory and, therefore, can not define 
an event horizon. Moreover, we expect (in analogy with the hydrogen atom 
model) the existence of bound states with discrete mass (energy) spectrum. 
Thus, the very process of quantum gravitational collapse should be quite 
different from the classical collapse.It seems reasonable to try to 
construct some simple quantum mechanical models in order  to understand, 
what the definition of quantum black hole could be. Our opinion is that we 
need to look for some special quantum state (or states) which have such a 
property as the absence of ``quantum hairs'' (loss of the initial data 
memory, appearance of some entropy and all that).

In a series of papers \cite{me} a quantum mechanical model of a 
self-gravitating spherically symmetric thin dust shell was constructed. 
Despite of simplicity, such a model reveals rather unexpected features. 

First of all, it appeared that bound states (in our cases we are 
dealing with the s-waves only) are described by two quantum numbers 
(instead of one number in conventional quantum mechanics). The 
spherically symmetric thin dust shell motion is described by three 
parameters, the Schwarzschild masses $m_{in}$ and $m_{out}=m$ 
inside and outside the shell, respectively, and the bare mass $M$. 
These parameters are connected to the above mentioned quantum 
numbers $n$ and $p$ by the following relations.
\begin{equation}
\label{n}
\frac{2 (\Delta m)^2 - M^2}{\sqrt{M^2 - (\Delta m)^2}} = 
2n \frac{m_{Pl}^2}{m},\ \ n -integer, 
\end{equation}
\begin{equation}
\label{p}
M^2 - (\Delta m)^2 = 2 (1 + 2p) m_{Pl}^2,\ \ p - integer.
\end{equation}

Here $ \Delta m = m-m_{in}$ is the total mass (energy) of the shell, 
and $m_{Pl}$ is the Planckian mass. The quantum number $n$ is 
reduced to the principal quantum number in the nonrelativistic 
Coulomb(or Kepler) problem in the corresponding limit 
($\Delta m \to M-0$). The appearance of the second quantum number 
$p$ is due to the nontrivial causal structure of the Schwarzschild 
space-time, where we have two asymptotically flat region connected 
by the so-called Einstein-Rosen bridge. If the shell in its 
(classical) motion goes through the asymptotically flat region 
on ``our side'' of the Einstein-Rosen bridge we will call this 
the black hole case, if it is on the ``other side'' - we will 
call this the wormhole case. In the first case 
$\frac{\Delta m}{M} > \frac{M}{4m}$ (and 
$\frac{\partial m}{\partial M}\mid_{n = const} > 0$ for quantum shells).
For the wormhole case $\frac{\Delta m}{M} < \frac{M}{4m}$ and 
$\frac{\partial m}{\partial M}\mid_{n = const} < 0$. 
Note that from Eqs.(\ref{n}) 
and (\ref{p}) it follows that there exists the minimal possible black 
hole (not wormhole!) mass. It corresponds to $ n = p = m_{min} = 0$ and 
equals $m_{min} = \sqrt{2} m_{Pl}$ for our particular model. 
The parameters of the shell $M$ and $\Delta m$ are imprinted in the 
wave functions of the bound states and can be, in principle, measured. 
Thus, we have ``quantum hairs'', and these states in no way can be 
considered as the quantum black hole states (zero entropy, no 
universality). Since the wave functions are nonzero everywhere we have 
a ``mixture'' of noncollapsing shells, collapsed shells and wormhole shells. 
What, then, the quantum black holes are? 

Before answering this question, let us have a look at another interesting 
feature of our model. Due to the nontrivial causal structure of the 
Schwarzschild space-time we can expect the existence of one 
quantization condition for the unbound states as well. And this is indeed 
the case. For the zero bare mass $M = 0$ we have a thin null shell that can 
be considered as a model for radiation. And the spectrum for such a radiation 
is discrete and given by 
\begin{equation}
\label{k}
3 (\delta m)^2 + 4 m \delta m = 4 k m_{Pl}^2, \ \ k -integer, 
\end{equation}
where $\delta m$ is the difference in the mass of the system before and 
after emission (change in Bondi's mass). The jump to the nearest level 
corresponds to $k = 1$. It is interesting that from the Eq.(\ref{k}) 
it follows that the gravitating system with the mass less than 
$m = 2/\sqrt{5}$ cannot radiate. The spectrum for the radiation is 
universal in the sense that it does not depend on the structure of the 
radiating system. Moreover, the energies of quanta of radiation 
(Eq.(\ref{k})) do not match the level  spacings of the shell 
(Eq.(\ref{k}) and \ref{k})). In order to satisfy the energy conservation 
law we have to suggest that together with every emitting quantum another 
shell(massive or massless) is created which collapses. Such a process 
increases the value of $m_{in}$ inside our (initially alone) shell, thus 
forming the internal structure of the (future) quantum black hole and 
creating its entropy.

Let us now look at the shells spectrum more attentively. We see that the 
Eq.(\ref{p}) depends only on the total mass (energy) of the shell 
$\Delta m$ and its bare mass $M$, while in the Eq.(\ref{n}) there 
enters also the inner Schwarzschild mass $m_{in}$ ($m=m_{in}+\Delta m$). 
But, there exists one special level with $n=0$ when such a dependence 
also disappears. If we take into account all the shell created during 
the quantum gravitational collapse and suppose that all of them are 
in the state with $n=0$ we obtain the quantum object that ``forgot'' 
all the past like the classical black hole ``forgets'' all the past 
during classical collapse. It seems that this is just what we would 
like to have for the quantum black holes. Note that in such a case 
\begin{equation}
\label{Mm}
\begin{array}{c}
M = \sqrt{2} m \\
\\
m = \sqrt{2(1 + 2p)} m_{Pl} .
\end{array}
\end{equation}
To support this point of view we construct some classical model 
which mimics the main features of the state with $n=0$. Let us 
consider the spherically symmetric distribution of perfect fluid 
that is initially at rest. Then, we can introduce the mass function 
$m(r)$ and the bare mass function $M(r)$ 
\begin{equation}
\label{mr}
m(r) = 4\pi \int \epsilon r^2 dr ,\\
\end{equation}
\begin{equation}
\label{Mr}
M(r) = 4\pi \int \epsilon \sqrt{- g_{11}} r^2 dr ,
\end{equation}
and solve the initial value problem 
\begin{equation}
\label{g}
g_{11} = -1 + \frac{2 G m(r)}{r} .
\end{equation}
Here $r$ is the radius (and, simultaneously, the radial coordinate), 
$g_{11}$ is the radial component of the metric tensor. In order to 
mimic the state with $n=0$ when nothing depends on the environment 
$m_{in}$ we have to put $m(r)=ar$ and, therefore, $\epsilon = 
\frac{a}{4\pi r^{2}}$. It is interesting to note that such a 
distribution is marginal between the distributions $\epsilon = 
\frac{a}{4\pi r^{2+\gamma}}$ with $\gamma >0$ (when near the origin 
there is a region where the initial state $\dot r=0$ is impossible, 
i.e., the black hole exists from the very beginning) those with 
$\gamma <0$ (when we can reach a surface with maximal area at large 
enough value of radius, i.e., we can construct a semi-closed world 
and have a wormhole case). In the case of pure dust distribution 
(no interaction except the gravitational one) the system would 
collapse. If initially the boundary of the system is $r_{0}$, 
then 
\begin{equation}
\label{m0}
\begin{array}{c}
m_0 = m(r_0) = a r_0 \\
\\
M_0 = M(r_0) = \frac{a r_0}{\sqrt{1 - 2Ga}} .
\end{array}
\end{equation}
Keeping $m_{0}$ constant and increasing $M_{0}$ we get 
\begin{equation}
\label{a}
\begin{array}{lcl}
a \longrightarrow \frac{1}{2 G} & \mbox{if} & M_0 \longrightarrow \infty\\
\\
r_0 = \frac{m_0}{a} & \longrightarrow & 2 G m_0 = r_g .
\end{array}
\end{equation}
That is, limit of infinite bare mass the initial boundary lies 
exactly at the gravitational radius of the system. But, our quantum 
system is stationary. Thus, to have the static situation in our 
classical analogue we need to introduce an effective pressure $p(r)$. 
Solving the Einstein`s equations we find that the pressure profile 
is the same as the energy density profile, but with different 
proportionality factor, namely 
\begin{equation}
\label{b}
\begin{array}{l}
p = \frac{b}{4\pi r^2} ,\\
\\
b = \frac{1}{G} (1 - 3Ga - \sqrt{1 - 2Ga}\sqrt{1 - 4Ga}) .
\end{array}
\end{equation}
We see that the coefficient $b$ is real only $a<a_{cr}=\frac{1}{4G}$. 
Beyond this value $b$ becomes a complex number. It is easy to show 
that at the critical value the speed of sound in our effective model 
reaches the speed of light. This means that even with the pressure 
we are unable prevent the system from collapsing! And, what is not 
seemed as a mere coincidence, at $a=a_{cr}$ 
\begin{equation}
\label{cr}
M_0 = \sqrt{2} m_0
\end{equation}              
exactly the same relation that we had before for the quantum state 
$n=0$!

\section*{Acknowledgments}

I am greatly indebted to Prof. Remo Ruffini whose kind invitation 
and financial support allowed me to attend the MG IX MM conference. 
I am grateful to Prof. Petr Hajicek for hospitality extended to me 
during my visit to the Institute for Theoretical Physics (Bern, 
Switzerland) and for fruitful discussions. I would like to thank 
the Russian Foundation for Basic Research  (grant 99-02-18524-a) for financial 
support.

\end{document}